\documentclass[prd,amsmath,amssymb]{revtex4}
\usepackage{psfrag}
\usepackage{graphicx}
\usepackage{graphics}
\usepackage{bm}
\usepackage{color}
\usepackage{amssymb}
\usepackage{amsmath}
\input{epsf.sty}

\newcommand{\nc}{\newcommand}
\newcommand{\nn}{\nonumber}
\nc{\ba}{\begin{eqnarray}}
\nc{\ea}{\end{eqnarray}}
\newcommand\be{\begin{equation}}
\newcommand\beq{\begin{equation}}
\newcommand\ee{\end{equation}}
\newcommand\eeq{\end{equation}}

\nc{\ga}{\gamma}
\nc{\tnu}{\bar{\nu}}
\nc{\tmu}{\bar{\mu}}
\nc{\tq}{\tilde{q}}

\nc{\x}{{\bf{x}}}
\newcommand{\gsim}{\raise.3ex\hbox{$>$\kern-.75em\lower1ex\hbox{$\sim$}}}
\newcommand{\lsim}{\raise.3ex\hbox{$<$\kern-.75em\lower1ex\hbox{$\sim$}}}
\nc{\gm}{\gamma }
\nc{\PIJ}{\tP_{,X_{IJ}}}
\nc{\hP}{\hat{P}}
\nc{\PX}{\tP_{,X}}
\nc{\half}{\frac{1}{2}}
\nc{\p}{\phi}
\newcommand{\dn}[2]{{\mathrm{d}^{{#1}}{{#2}}}}
\nc{\s}{\sigma}

\newcommand\bk{\boldsymbol{k}}
\nc\pp{\psi}
\nc\bfphi{{\bf \phi}}
\def\a{\alpha}
\def\b{\beta}
\def\H{{\cal F}}
\def\tP{{\tilde P}}
\def\tX{{\tilde X}}
\def\tG{{\tilde G}}
\def\e{e}

\def\R{{\cal R}}
\def\N{{\cal N}}
\def\A{{\cal A}}
\def\M{A}

\def\Tdot#1{{{#1}^{\hbox{.}}}}
\def\Tddot#1{{{#1}^{\hbox{..}}}}

\begin{document}
\title{Primordial perturbations and non-Gaussianities in \\ DBI and general multi-field inflation}

\author{David Langlois$^{1,2}$, S\'ebastien Renaux-Petel$^1$, Dani\`ele A.~Steer$^{1,3}$ and Takahiro Tanaka$^4$}
\affiliation{$^1$APC (UMR 7164, CNRS, Universit\'e Paris 7), 10 rue Alice Domon et L\'eonie Duquet, 75205 Paris Cedex 13, France}
\affiliation{$^2$IAP, 98bis Boulevard Arago, 75014 Paris, France}
\affiliation{$^3$CERN Physics Department, Theory Division, CH-1211 Geneva 23, Switzerland}
\affiliation{$^4$Department of Physics, Kyoto University, Kyoto, 606-8502, Japan}

\date{\today}

\begin{abstract}

We study cosmological perturbations in general inflation models with multiple scalar fields and arbitrary kinetic terms, with special emphasis on the multi-field extension of Dirac-Born-Infeld (DBI) inflation.  We compute the second-order action governing the dynamics of linear perturbations in the most general case. Specializing to DBI, we show that the adiabatic and entropy modes propagate with a {\it common} effective sound speed  and are thus amplified at sound horizon crossing.  In the small sound speed limit, we find that the amplitude of the entropy modes is much higher than that of the adiabatic modes. We also derive,
in the general case, the third-order action which is useful for studying primordial non-Gaussianities generated during inflation. In the DBI case,  we compute the dominant contributions to non-Gaussianities, which depend on both the adiabatic  and entropy modes. 
\end{abstract}
\maketitle


\section{Introduction}

While inflation has become a standard paradigm with which to describe the physics of the very early universe, the nature of the field(s) responsible for inflation remains an open question. The last few years have seen an intensive effort devoted to trying to connect string theory and inflation (for recent reviews, see e.g.~\cite{McAllister:2007bg,Burgess:2007pz,Kallosh:2007ig,Cline:2006hu,HenryTye:2006uv}), with the hope that future cosmological observations, in particular of the CMB anisotropies, could detect some specific stringy signatures.

 Of  particular interest are scenarios based on the motion of a D-brane in a higher-dimensional spacetime. Since 
the dynamics of a D-brane is described by the Dirac-Born-Infeld (DBI) action, characterized by a non-standard kinetic term, inflation can occur with steep potentials, in contrast with usual slow-roll inflation. In this sense, this 
scenario, called  DBI inflation \cite{st03,ast04,Chen:2004gc,Chen:2005ad},  belongs to the more general class of \textit{k}-inflation models
\cite{ArmendarizPicon:1999rj,Garriga:1999vw} characterized by a Lagrangian of the form $P(X,\phi)$, where 
$X=-\partial_\mu\phi \partial^\mu\phi/2$.

In DBI inflation, the effective four-dimensional scalar field corresponds to the radial position of a brane in a higher dimensional warped conical geometry. For simplicity, the other possible degrees of freedom of the brane, namely the angular coordinates, are usually assumed to be frozen. Relaxing this assumption and allowing the brane to move in the angular directions (see e.g \cite{Kehagias:1999vr}-\cite{Easson:2007fz}) leads to a multi-field scenario, since each brane coordinate in the extra dimensions gives rise to a scalar field from the effective four-dimensional point of view. 

 Beyond the multi-field extension of DBI inflation, it is interesting to study, in the spirit of \textit{k}-inflation, very general multi-field Lagrangians of the form 
\beq
P=P\left(X^{IJ},\phi^K \right),
\label{P_XIJ}
\ee
with 
 \be
 \label{def-XIJ}
X^{IJ} \equiv -\half \partial_\mu \p^I  \partial^\mu \p^J \, ,
\eeq
where $I=1,\ldots,N$ labels the multiple fields.  (In the following we will adopt the implicit summation convention on both field and space-time indices.)

A more restrictive class of models consists of Lagrangians that depend on the global kinetic term $X=G_{IJ}X^{IJ}$
where the functions $G_{IJ}(\phi^K)$ are the components of some metric defined in field space \cite{ds}. While this simpler class of Lagrangians is enough to describe the {\it homogeneous} dynamics of multi-field DBI inflation, it turns out that the full {\it inhomogeneous} dynamics {\it cannot} be described by such a Lagrangian, 
as we pointed out in \cite{Langlois:2008wt} and show below in more detail.

The purpose of the present work is thus two-fold. Our first aim is to derive the equations governing cosmological perturbations in the generalized models of the form given in Eq.~(\ref{P_XIJ}).
Our second aim is to apply this general formalism to the multi-field extension of the DBI scenario.

The structure of this paper is the following. In the next section, we first consider the multi-field DBI action
which motivates our subsequent study of the general formalism. In section \ref{sec:3} we derive, in the general case, the field equations for the metric and for the scalar fields, after which we specialize to the homogeneous background. 
Section \ref{sec:4} is devoted to the dynamics of the linear perturbations in the general case: we derive the second-order action and analyse the resulting equations of motion for the perturbations. 
We then focus, in section \ref{sec:5}, on the specific example of the DBI action: we show that the adiabatic and entropy modes propagate with the same speed of sound $c_s$ and we compute the second-order action for linear perturbations.   For two-field DBI inflation we also compute the power
spectra of the adiabatic and entropy modes.
Finally in section \ref{sec:6} we discuss non-Gaussianities.  We first derive, in the general case,
the third-order action for perturbations. We then limit our analysis to two-field DBI models, for which we compute the main contribution to non-Gaussianity in the limit of small $c_s$. 
We summarize our main results in the last section.

\section{The multi-field DBI action}
\label{sec:2}

In this section we motivate our reasons for looking at Lagrangians of the general form $P(X^{IJ},\phi^K)$ by showing, in particular, that multi-field DBI inflation is described by a Lagrangian of this form. We also discuss the properties of the higher order terms in derivatives which appear in the DBI Lagrangian.

Consider a D3-brane with tension $T_3$ evolving in a 10-dimensional geometry described by the metric
\be
ds^2 = h^{-1/2}(y^K)\,g_{\mu \nu}dx^\mu dx^\nu + h^{1/2}(y^K)\, G_{IJ}(y^K)\, dy^I dy^J \equiv H_{AB} dY^A dY^B
\ee
with coordinates $Y^A=\left\{x^\mu, y^I\right\}$, 
where $\mu=0,\ldots 3$ and $I=1,\ldots, 6$ (the label $I$ has been chosen in this way as below it will label the multiple effective scalar fields).  The kinetic part of the DBI action, 
\beq
\label{L}
L =- T_3\sqrt{-\det{\gamma_{\mu \nu}}}
\eeq
depends on the determinant of the induced metric on the 3-brane,
\be
\gamma_{\mu \nu} =H_{AB} \partial_\mu Y_{\rm (b)}^A \partial_\nu Y_{\rm (b)}^B 
\ee
where the  brane embedding is defined by the functions  $Y_{\rm (b)}^A(x^\mu)$, with the $x^\mu$ being the spacetime coordinates on the brane. In our case, they coincide with the first four bulk coordinates. On writing
$Y_{\rm (b)}^A = (x^\mu, \varphi^I(x^{\mu}))$, we find
\be
\gamma_{\mu \nu} = h^{-1/2} \left( g_{\mu \nu}  + h \, G_{IJ} \partial_\mu \varphi^I \partial_\nu \varphi^J \right)
\label{DBIzero}\, ,
\ee
which after substitution into (\ref{L}) implies
\beq
L=-T_3 h^{-1} \sqrt{-g} \, \sqrt{\det(\delta^{\mu}_{\nu}+h \, G_{IJ}\partial^{\mu} \varphi^I \partial_{\nu} \varphi^J )}\, .
\eeq
Finally, in order to absorb the brane tension $T_3$, it is convenient to rescale in the following way:
\be
f= \frac{h}{T_3} \; , \qquad \phi^I = \sqrt{T_3}\varphi^I.
\label{redef}
\ee
As a result, in the following, we consider the DBI Lagrangian
\beq
P= -\frac{1}{f(\bfphi^I)}\left(\sqrt{{\cal D}}-1\right) -V(\bfphi^I)
\label{DD}
\eeq
where
\beq
{\cal D}=  \det(\delta^{\mu}_{\nu}+f \, G_{IJ}\partial^{\mu} \p^I \partial_{\nu} \p^J )\,,
\label{Ddef}
\ee
and where we have also included  potential terms, which arise from the brane's interactions with bulk fields or other branes. From now on we let $I=1,\ldots,N$.

One can express the above Lagrangian in (\ref{DD}) explicitly in terms of the $X^{IJ}$ defined in (\ref{def-XIJ}), 
by rewriting ${\cal D}$, which is the determinant of a $4\times 4$ matrix,  as the determinant 
of an $N \times N$ matrix:
\beq
{\cal D}=\det(\delta_{I}^{\, J}-2f X_{I}^{\, J}) \label{det}
\ee
where
\beq
X_I^J=G_{IK}X^{KJ}.
\label{def-XIJ2}
\eeq
Throughout this paper field indices will always be raised and lowered with the `field metric' $G_{IJ}=G_{IJ}(\phi^K)$.
The equality between the expressions (\ref{Ddef}) and (\ref{det}) for the determinant
  can be proved by using  the identity $\det(\mathbf{Id}+{\mathbf{\alpha}})=\exp[{\rm Tr}({\rm ln} (\mathbf{Id}+{\mathbf{\alpha}}))]$ for the matrix $\mathbf{\alpha}$ of components $\alpha^\mu_\nu = f G_{IJ} \partial^\mu \phi^I \partial_\nu \phi^J$.  Indeed from Eq.~(\ref{Ddef}) we have
\be
{\cal D}=\exp\left[{\rm Tr}(\alpha)-\half {\rm Tr}(\alpha^2)+\frac{1}{3}{\rm Tr}(\alpha^3)+...\right]\, ,
\label{intermediate}
\ee
and on noting that
\be
{\rm Tr}(\alpha^n)={\rm Tr}\left[(-2f {\cal X})^n\right]\, 
\label{remark}
\ee
where ${\cal X}$ represents the matrix of components $X_{I}^{\, J}$, one obtains the expression given in Eq.~(\ref{det}).

Another very useful expression for ${\cal D}$ can be obtained by computing directly the determinant in  Eq.~(\ref{Ddef}). As we show in Appendix 1, this yields 
\beq
\label{def_explicit}
{\cal D}=1-2f G_{IJ}X^{IJ}+4f^2 X^{[I}_IX_J^{J]} -8f^3 X^{[I}_IX_J^{J} X_K^{K]}+16f^4 X^{[I}_IX_J^{J} X_K^{K}X_L^{L]},
\eeq
where the brackets denote antisymmetrisation on the field indices. We note that,  in four  spacetime dimensions, Eq.~(\ref{def_explicit}) is automatically truncated at order $f^4$ even if the number of scalar fields is larger than four (see Appendix 1).
To use shorter notations, we will rewrite the above equation as
\beq
{\cal D}=1-2f \tX,
\label{short}
\eeq
with 
\ba
\tilde{X} &\equiv & X +  \H(X^{IJ},\phi^K),
\\
X &\equiv&   G_{IJ}X^{IJ}
\label{Xdef}
\ea
and where $\H(X^{IJ},\phi^K)$ can be read from Eq.~(\ref{def_explicit}):
\beq
\H(X^{IJ},\phi^K) = -2f X^{[I}_IX_J^{J]} +4 f^2 X^{[I}_IX_J^{J} X_K^{K]}-8f^3 X^{[I}_IX_J^{J} X_K^{K}X_L^{L]}.
\label{Fdef}
\eeq

For a single field, $I=1$, it is straightforward to see that $\H$ vanishes so that the determinant takes the familiar form ${\cal D}=1+f\partial_\mu\phi\partial^\mu\phi$ (for $G_{11}=1$). Similarly, for a multi-field
 {\it homogeneous} configuration in which the scalar fields depend only on time and ${X}^{IJ}=\dot\phi^I\dot\phi^J/2$, one again finds $\H=0$ because of the antisymmetrisation on field indices in Eq.~(\ref{Fdef}).  Thus in this case the determinant  ${\cal D}$ reduces to 
   \beq
\bar{{\cal D}}=1- f\, G_{IJ} \dot \p^I \dot \p^J.
\eeq
(In the following a bar denotes homogeneous background quantities.)

 However, for {\it multiple inhomogeneous} scalar fields, the terms in $\H$, which are higher order in gradients and have not been considered in previous works, 
do not vanish: we will show later in this paper that they drastically change the behaviour of perturbations. Furthermore, since they vanish in the homogeneous background, we expect them  to modify only the terms in the perturbation equations which contain spatial derivatives. From this discussion we therefore see explicitly that the multi-field DBI action does not depend only on $X=G_{IJ}X^{IJ}$ (as has been assumed in recent works on multi-field DBI inflation \cite{Easson:2007dh,Huang:2007hh}), but requires a general description of the form $P = P(X^{IJ},\phi^K)$.

After this digression on the specific form of the multi-field DBI Lagrangian, in the following section we return to 
 the general Lagrangian given in Eq.~(\ref{P_XIJ}).

\section{Field equations}
\label{sec:3}

We begin this section by deriving the equations of motion for the general action
\beq
\label{action}
S = \frac{1}{2} \int {\rm d}^4 x \sqrt{-g}\left[ {}^{(4)}{{ R}}   +  2 P(X^{IJ},\phi^K)\right],
\eeq
where we have set $8\pi G=1$.
The energy-momentum tensor can be obtained by varying $P$ with respect to the metric, and is given by 
\be
T^{\mu \nu} = P g^{\mu \nu} + P_{<IJ>}  \partial^\mu \phi^I \partial^\nu \phi^J\,,
\label{theend}
\ee
where we have defined 
\be
P_{<IJ>}  \equiv  \frac{1}{2} \left( \frac{\partial P}{\partial X^{IJ}}  + \frac{\partial P}{\partial X^{JI}}  \right) = P_{<JI>} \, .
\label{lafinpeutetre}
\ee
We use this symmetrized derivative of the Lagrangian $P$ with respect to $X^{IJ}$ for the following reason: since $X^{IJ}$ is symmetric in $I$ and $J$, the explicit dependence of $P$ on say $X^{12}$ and $X^{21}$ can vary depending on the chosen convention and the above definition avoids any ambiguity. The same notation will apply to the derivative of any arbitrary function which depends on $X^{IJ}$.

The equations of motion for the scalar fields follow from the variation of the action in Eq.~(\ref{action}) with respect to each scalar field. One finds 
\be
\label{KG0}
\frac{1}{\sqrt{-g}} \partial_{\mu} \left( \sqrt{-g} \, P_{<IJ>} \partial^\mu \phi^J \right) + P_{,I} = 0\,,
\ee
where $P_{,I}$ denotes a partial derivative with respect to $\phi^I$.

Now we consider a spatially flat  FLRW (Friedmann-Lema\^itre-Robertson-Walker) geometry with metric
\be
ds^2 = -dt^2 + a(t)^2 d{\bf x}^2\, ,
\label{bck}
\ee
where $t$ is cosmic time.  Then the kinetic terms defined in Eq.~(\ref{def-XIJ}) reduce to
\be
{X}^{IJ} =  \frac{1}{2}\dot{\phi}^I  \dot{\phi}^J\,,
\ee
where a dot denotes a derivative with respect to $t$.
From Eq.~(\ref{theend}), the energy density can be expressed as 
\be
\label{rho}
\rho = 2  P_{<IJ>}  X^{IJ} - P
\ee
while the pressure is simply $P$, and the Friedmann
 equations are given by
\be
H^2 = \frac{1}{3} \left(2  P_{<IJ>}  X^{IJ} - P \right)\, , \qquad
\dot{H} = - X^{IJ} P_{<IJ>} \,.
\label{Fried}
\eeq
The equations of motion for the scalar fields Eq.~(\ref{KG0}) reduce to 
\be
a^{-3} \frac{{\rm d}}{{\rm dt}} \left( a^3 P_{<IJ>} \dot{\phi}^J \right) = P_{,I} \, .
\label{compact KG}
\ee
On calculating the time derivative and taking into account the terms in $\ddot{\phi}^J$ contained in $\frac{{\rm d}}{{\rm dt}}{P}_{<IJ>} $, the above equation can be rewritten as
\be
\left( P_{<IJ>} + P_{<IL>,<JK>} \dot{\phi}^L \dot{\phi}^K  \right) \ddot{\phi}^J+ 
 \left( 3 H P_{<IJ>} + {P}_{<IJ>,K} \dot{\phi}^K\right)\dot{\phi}^J - P_{,I} = 0
\label{KG}
\ee
where, in analogy with Eq.~(\ref{lafinpeutetre}), we have defined 
\be
 P_{<IJ>,<KL>} \equiv   \frac{1}{2} \left( \frac{\partial P_{<IJ>}}{\partial X^{KL}}  + \frac{\partial P_{<IJ>}}{\partial X^{LK}}  \right)= P_{<KL>,<IJ>}.
\ee

Finally we end this section by noting that were
the Lagrangian to depend  on the $X^{IJ}$ {\it only} through $X=G_{IJ}X^{IJ}$, then one would define 
$\tilde{P}(X,\phi^K)=P(X^{IJ},\phi^K)$ so that $P_{<IJ>} = \tilde{P}_{,X} G_{IJ}$. In that case all
the above expressions would reduce to those of \cite{ds}.

\section{Linear perturbations in the general case}
\label{sec:4}

In this section, we derive the second-order action governing the dynamics of the linear perturbations for the general action given in Eq.~(\ref{action}). As in \cite{Maldacena:2002vr}, we use the ADM approach,   and our calculations are very similar to those of \cite{ds}, except that we work with the quantities $X^{IJ}$ rather than $X$.

Starting from the metric in the ADM form
\beq
ds^2=-N^2 dt^2 +h_{ij} (dx^i+N^i dt)(dx^j+N^j dt)\,
\label{metric}
\eeq
where $N$ is the lapse and $N^i$ the shift,
the full action becomes
\ba
\label{action-ADM}
S &=& \frac{1}{2} \int {\rm d}^4x \sqrt{-g} \left(^{(4)}{{R}} + 2 P \right)
= \frac{1}{2} \int {\rm d}t \, {\rm d}^3x  \sqrt{h} \left[ N \; ^{(3)}{{R}}  + \frac{1}{N} (-E^2 + E_{ij}E^{ij} ) + 2 N P \right]
\ea
where $^{(3)}{{R}}$ is the scalar curvature of the spatial metric $h_{ij}$ with $h$ its determinant, and  the symmetric tensor $E_{ij}$, defined by 
\beq
E_{ij}=\half \dot{h}_{ij}-D_{(i}N_{j)}
\label{Eij}
\eeq
is proportional to the extrinsic curvature of the spatial slices ($D_i$ denotes the spatial covariant derivative associated with 
the spatial metric $h_{ij}$).

The function $P=P(X^{IJ},\phi^K)$ in Eq.~(\ref{action-ADM}) depends on the kinetic quantities $X^{IJ}$, which can be decomposed as
\be
X^{IJ} = \frac{1}{2N^2} v^I v^J - \frac{1}{2}h^{ij} \partial_i \phi^I \partial_j \phi^J
\ee
with
\be
v^J  = \dot{\phi}^J - N^i \partial_i \phi^J.
\ee
We now work in the flat gauge so that spatial sections are flat, $h_{ij} = a^2(t)\delta_{ij}$ and $^{(3)}{{R}}=0$.
The Hamiltonian and momentum constraints, which follow from the variation of (\ref{action-ADM}) with 
respect to the lapse and the shift are, respectively,
\ba
2({N^2}P -  P_{<IJ>}v^I v^J) + E^2 - E_{ij}E^{ij} &=&0\,,
\\
 D_j \left[ \frac{1}{N} \left(E^j_i - E \delta^{j}_i \right) \right] &=& 
 \frac{1}{N}   P_{<IJ>}v^{I}  \partial_i \phi^{J} \,. 
 \label{momentum_constraint}
\ea
To zeroth order (background), the Hamiltonian constraint simply gives the first Friedmann equation in Eq.~(\ref{Fried}), while  the momentum constraint vanishes identically.

At first order in the scalar perturbations, we write
\be
N = 1 + \delta N, \qquad N_i = \partial_i \pp, \qquad \phi^I = \bar{\phi}^I(t) + Q^I(t,{\bf x})
\label{lapse-shift}
\ee
where, from now on and when there is no ambiguity, we promptly drop the bars on all the unperturbed fields.  
Note that $\psi$ is related to the standard Bardeen potential $\Psi$ by
\be
\Psi = - H \psi \, .
\label{Bardeen}
\ee
At linear order,
the momentum constraint (\ref{momentum_constraint}) gives
\be
\delta N = \frac{1}{2H}  P_{<IJ>} \dot{\phi}^{I} Q^{J}.
\label{dN}
\ee
The Hamiltonian constraint is  more complicated, but a straightforward calculation yields
\be
-2H \left(\frac{\partial^2 \pp}{a^{2} }\right) = 2A \delta N +B_{IJ} \dot{\phi}^{J} \dot{Q}^{I} + C_I Q^I
\label{supercomplicated}
\ee
with
\ba
&& A  = P_{<IJ>} X^{IJ} - P - 2 X^{IJ} X^{KL} P_{<IJ>,<KL>},
\nn
\\
&& B_{IJ} = P_{<IJ>} + 2 X^{KL} P_{<IJ>,<KL>}\, ,
\nn
\\
&& C_I = - P_{,I} + 2  P_{<KL>,I}X^{KL}.
\ea
Actually, this explicit expression for $\pp$ is not necessary in order to derive the second-order action, as the terms involving $\pp$ (coming from the matter and gravitational parts of the action) cancel each other.   The scalar field perturbations are related to a useful geometrical quantity, namely the comoving curvature perturbation $\R$ (see e.g.~\cite{mfb} and \cite{Langlois:2004de}).  On using the standard definition of $\R$ which combines the metric perturbations with the perturbations of the momentum density for the Lagrangian $P=P(X^{IJ},\phi^K)$, one obtains   
\be
\R = \left( \frac{H}{2 P_{<IJ>}X^{IJ}} \right) P_{<KL>} \dot{\phi}^K Q^L.
\label{calRdef}
\ee

After these preliminary steps, one can now expand the action (\ref{action-ADM}) up to second order in the linear perturbations $\delta N$, $\psi$ and $Q^I$.  As mentioned earlier, the terms involving $\psi$ cancel each other.  On reexpressing $\delta N$ in terms of the $Q^I$
using the constraint (\ref{dN}), one obtains, after a long but straightforward calculation,
\ba
S_{(2)} &=& \frac{1}{2} \int {\rm d}t \, {\rm d}^3x \, a^3 \left[ 
\left(P_{<IJ>} + 2 P_{<MJ>,<IK>}X^{MK}\right) \dot{Q}^{I}\dot{Q}^{J}  - P_{<IJ>} h^{ij} \partial_iQ^I \partial_jQ^J \right.
\nn
\\
&& \qquad \qquad \qquad \qquad \left. - {\cal M}_{KL}Q^K Q^L 
 + 2 \,\Omega_{KI}Q^K \dot{Q}^I  \right] \qquad
 \label{S2}
\ea
where the mass matrix is
\ba
{\cal M}_{KL} &=& - P_{,KL} + 3 X^{MN} P_{<NK>} P_{<ML>} 
+ \frac{1}{H} P_{<NL>}\dot{\phi}^N \left[ 2 P_{<IJ>,K} X^{IJ} - P_{,K} \right]
\nn
\\
& - & \frac{1}{H^2} X^{MN} P_{<NK>}P_{<ML>} \left[ X^{IJ} P_{<IJ>} + 2 
P_{<IJ>,<AB>} X^{IJ}X^{AB} \right] 
\nn
\\
& - & { \frac{1}{a^3}\frac{{\rm d}}{{\rm dt}} } \left( \frac{a^3}{H} P_{<AK>} P_{<LJ>} X^{AJ} \right)
\label{masssq}
\ea
and the mixing matrix is
\be
\Omega_{KI} = \dot{\phi}^J P_{<IJ>,K} -
 \frac{2}{H} P_{<LK>} 
 P_{<MJ>,<NI>} X^{LN}X^{MJ}\,.
\label{damping}
\ee

On denoting the coefficient of the kinetic parts in Eq.~(\ref{S2}) by
\be
K_{IJ} \equiv  P_{<IJ>} + 2 P_{<MJ>,<IK>}X^{MK} ,
\label{Kdef}
\ee
we find that the equations of motion for the $Q^I$ (in Fourier space) are
\be
{ K_{IJ}\ddot{Q}^J +  \frac{k^2}{a^2} P_{<IJ>}Q^J + \left(\dot{K}_{IJ} +3 H K_{IJ}+ \Omega_{JI} - \Omega_{IJ}\right) \dot{Q}^J}+\left( \dot{\Omega}_{KI} + {\cal M}_{IK} +3 H \Omega_{KI} \right) Q^K= 0\,.
\ee
The propagation velocities can be deduced from the structure of the first two terms in the above equation. 
On assuming that $K_{IJ}$ is invertible, the sound speeds correspond to the eigenvalues of the matrix of components
$(K^{-1})^{IL}P_{<LJ>}$ (recall that these are background quantities).

For a {\it single scalar field}, $P_{<IJ>}$ reduces to $P_{,X}$, and it is easy to see that the kinetic coefficient in Eq.~(\ref{Kdef}) is simply $K=P_{,X} +2X P_{,XX}$ which can be identified with $\rho_{,X}$ according to the relation (\ref{rho}). Hence one recovers the familiar result \cite{Garriga:1999vw} that the effective speed of sound is given by
\beq
\label{cs}
c_s^2=\frac{P_{,X}}{\rho_{,X}}=\frac{P_{,X}}{P_{,X}+2XP_{,XX}} 
\qquad \left({\rm single \ scalar \ field}\right).
\eeq
For {\it multiple fields} and in the particular case where the Lagrangian is a function of $X=G_{IJ} X^{IJ}$, i.e. 
$P=P(X,\phi^K)$, it has been shown in \cite{ds} (see also \cite{Easson:2007dh,Huang:2007hh}) that the propagation matrix $(K^{-1})^{IL}P_{<LJ>}$ becomes {\it anisotropic}: the perturbations along the field space trajectory propagate with an effective speed of sound $c_s$, defined as in the single field case above (\ref{cs}), whereas the perturbations orthogonal to the background trajectory propagate at the speed of light. In the next section, we examine what happens in multi-field DBI inflation for which $P=P(X^{IJ},\phi^K)$.

\section{Linear perturbations in  DBI inflation}
\label{sec:5}

We now focus on linear perturbations in the specific case of multi-field DBI inflation for which the Lagrangian was derived in section 2:
\be
P(X^{IJ},\phi^K)= -\frac{1}{f(\bfphi^I)}\left(\sqrt{{\cal D}}-1\right) -V(\bfphi^I),
\label{DD1}
\ee
where ${{\cal D}}$ is given in Eq.~(\ref{det}) or Eq.~(\ref{def_explicit}).
According to Eq.~(\ref{short}), this Lagrangian can also be seen as a function of $\tilde{X}$, and it will be convenient later to use $\tP$ defined by
\beq
 P(X^{IJ},\phi^K) \equiv  \tilde{P}(\tilde{X},\phi^K) = -\frac{1}{f(\bfphi^I)} \left(\sqrt{1-2f(\bfphi^I)\tilde{X}}-1 \right) - V(\bfphi^I).
\label{Ptilde}
\eeq

\subsection{Propagation speed}

Before considering the full dynamics of the linear perturbations, it is instructive to concentrate on their propagation speed. According to the general analysis given in the previous section, we simply need to calculate  $P_{<IJ>}$ as well as the matrix  $K_{IJ}$, defined in Eq.~(\ref{Kdef}).

To do so, it is convenient to use the form of the determinant given in Eq.~(\ref{det}), namely
\beq
\label{det2}
{\cal D}=\det(M), \qquad M_{I}^{\, J}\equiv \delta_{I}^{\, J}-2f G_{IK} X^{KJ}  \,.
\ee
In the homogeneous background, the components of the matrix $M_{I}^{\, J}$ reduce to 
\beq
\label{Mbar}
\bar{M}_{I}^{\, J}=\delta_{I}^{\, J}-2f X e_I e^J,
\eeq
when expressed in terms of $X=G_{IJ}X^{IJ}$ and of the unit
vector in field space
\beq
{ \e^I \equiv \frac{\dot{\phi}^I}{\sqrt{2X}}, \quad e_I\equiv G_{IJ}e^J.}
\eeq
We will shortly need the inverse of the background matrix $\bar{M}$, denoted by $\tilde{G}$. Its  components
 are given by 
 \beq
\label{Gtilde0}
\tilde{G}_{I}^{\, J}=\delta_{I}^{\, J}+\frac{2fX}{1-2fX} e_I e^J=\; \perp_{I}^{\, J}+\frac{1}{1-2fX} e_I e^J
\eeq
where, in the second equality, we have introduced  the projector orthogonal to the vector $e^I$,
\beq
\perp_{I}^{\, J}=\delta_{I}^{\, J}-e_Ie^J\, .
\eeq
Let us also define
\be
c_s = \sqrt{1-2fX} ={\bar {\cal D}}^{1/2}
\label{csDBI}
\ee
which, as we show below, is the propagation speed for {\it all} perturbations.  Note that this definition coincides with that given in Eq.~(\ref{cs}) for a single scalar field (on replacing $P(X,\phi^I)$ by $\tilde{P}(\tilde{X},\phi^I)$ given in Eq.~(\ref{Ptilde})), and that $\tilde{G}$ given in Eq.~(\ref{Gtilde0}) can be rewritten as
\beq
\label{Gtilde} 
\tilde{G}_{I}^{\, J}= \, \perp_{I}^{\, J}+\frac{1}{c_s^2} e_I e^J.
\eeq

Let us now compute $P_{<IJ>}$.  The identity $\delta \det(M)=\det({\bar M}) {\bar M}^{-1}\delta { M}$ implies, using Eq.~(\ref{csDBI}), that
\beq
{\cal D}_{<IJ>}
=-2 f c_s^2\tG_{IJ}, 
\eeq
where $\tilde{G}_{IJ}=\tilde{G}_{I}^{\, K} G_{KJ}$. 
It then follows from Eq.~(\ref{DD1})  that
\beq
P_{<IJ>}=-\frac{1}{2fc_s} {{\cal D}_{<IJ>}}= c_s\tG_{IJ},
\label{PIJ}
\eeq
where all quantities are evaluated on the background.
For the matrix $K_{IJ}$, one needs the second derivative of $P$ with respect with $X^{IJ}$. On using the second derivative of the determinant ${\cal D}$, which can be deduced from Appendix 1, it is straightforward to obtain 
\be
 {P}_{<IK>,<JL>} =fc_s \left( \tilde G_{IL} \tilde G_{KJ}+ \tilde G_{IJ} \tilde G_{KL}- \tilde G_{IK} \tilde G_{JL}\right)\,.
\label{P(IJ)(KL)}
\ee
By noting that $X^{KL}\tG_{KL} = X/c_s^2$, the above equation together with Eq.~(\ref{PIJ}) leads to
\be
{ K_{IJ}\equiv {P}_{<IJ>}+2{P}_{<IK>,<JL>}X^{KL}=\frac{1}{c_s}\tilde G_{IJ}=\frac{1}{c_s^2}{P}_{<IJ>}\,.}
\label{K}
\ee
Hence we obtain the remarkable result that the propagation matrix is proportional to the identity matrix and that all perturbations propagate at the same speed, namely the effective sound speed $c_s$ defined in Eq.~(\ref{csDBI}).

Intuitively one can understand this result as follows.  Let us return to the DBI action in terms of the embedding of a brane in a higher dimensional spacetime, as discussed in Section \ref{sec:2}.  The perturbations we have considered above can be seen as fluctuations of the brane position in the higher dimensional background.  Since the brane action is the world-sheet volume, its fluctuations propagate at the speed of light from the higher dimensional point of view.  From a 4-dimensional point of view, this translates into the speed of sound $c_s$ as a consequence of time-dilation between the bulk time coordinate and the brane proper-time (the Lorentz factor is $1/c_s$).

\subsection{Second-order action for the perturbations}

We now turn to the full dynamics of linear perturbations in multi-field DBI inflation.  In the second-order order action (\ref{S2}), the mass and mixing terms could be determined by explicit substitution of $P$ given in Eq.~(\ref{DD1}).  Here, however, we follow a more direct route and extend the results 
of \cite{ds} (which were obtained for Lagrangians  depending only on $X$).  
To do so, we use $\tilde{P}(\tilde{X},\phi^I)$ defined in Eq.~(\ref{Ptilde}) and simply identify the new terms which appear because the DBI Lagrangian depends on $\tX$ rather than $X$.

Computation of the first and second order variations of $\tX$ gives (see Appendix 2)
\beq
\label{delta_var}
\delta^{(1)}\tX=\delta^{(1)} X, \quad \delta^{(2)}\tX=\delta^{(2)}X +{fX}\perp_{IJ} h^{ij} \partial_i Q^I\partial_j Q^J.
\eeq
With respect to the second order action of \cite{ds}, the extra term in Eq.~(\ref{delta_var}) {\it only} modifies the spatial gradient term, while
the rest of that action is unchanged.  Hence, as in \cite{ds}, we can rewrite the action in terms of covariant derivatives $ \mathcal{D}_I$ defined with respect to the the field space metric $G_{IJ}$.  This gives
\begin{eqnarray}
S_{(2)}&=& \half \int {\rm d}t \,\dn{3}{x}\,    a^3 \left[ \frac{1}{c_s} \left(\tG_{IJ}
 \mathcal{D}_t Q^I \mathcal{D}_t Q^J - c_s^2 \tG_{IJ}h^{ij} \partial_i Q^I \partial_j Q^J \right)
  - {\tilde{\cal M}}_{IJ}Q^I Q^J + 2  \frac{f_J X}{c_s^{3}}\dot \p_I Q^J \mathcal{D}_t Q^I  \right]\,,
\label{2d-order-action}
\end{eqnarray}
where we have substituted $\tP_{,\tX}=1/c_s$ and $ \tP_{,\tX J}  ={f_J X}/{c_s^{3}}$ into the expression of \cite{ds}, and introduced the time covariant derivative $\mathcal{D}_t Q^{I} \equiv \dot{Q}^I + \Gamma^{I}_{JK} \dot{\phi}^J Q^{K}$  where $\Gamma^{I}_{JK} $ is the Christoffel symbol constructed from $G_{IJ}$ (and $\mathcal{R}_{IKLJ}$ will denote the corresponding Riemann tensor).
Finally the mass matrix which appears above, and which differs from ${\cal M}_{IJ}$ in Eq.~(\ref{masssq}), is
\begin{eqnarray}
{\tilde{\cal M}}_{IJ} &=& -\mathcal{D}_I \mathcal{D}_J \tP - \tP_{,\tX} \mathcal{R}_{IKLJ}\dot \p^K \dot
\p^L+\frac{X \tP_{,\tX}}{H} (\tP_{,\tX J}\dot \p_I+\tP_{,\tX I}\dot \p_J)+ \frac{X \tP_{,\tX}^3}{2 H^2}\left(1-\frac{1}{c_s^2}\right)\dot \p_I \dot \p_J
 \nonumber\\
 &&~~{} -\frac{1}{a^3}\mathcal{D}_t\left[\frac{a^3}{2H}\tP_{,\tX}^2\left(1+\frac{1}{c_s^2}\right)\dot \p_
I \dot \p_J\right]  
\nonumber
\\
&=& \mathcal{D}_I \mathcal{D}_J V-\frac{(1-c_s)^2}{2 c_s }\frac{ \mathcal{D}_I \mathcal{D}_J f}{f^2}-\frac{(1-c_s)^3(1+3 c_s)}{4 c_s^3}\frac{\mathcal{D}_I f\, \mathcal{D}_Jf}{f^3}
+2 \dot{H} \mathcal{R}_{IKLJ}e^K e^L
+\frac{(1-c_s^2)^2}{2 c_s^4 f^2 H}f_{,(I} \dot{\p}_{J)}
 \nonumber\\
 &&~~{} + \frac{\dot{H}}{2H^2c_s^4}\left(1-c_s^2\right)\dot \p_I \dot \p_J
-\frac{1}{a^3}\mathcal{D}_t\left[\frac{a^3}{2H c_s^4}\left(1+c_s^2 \right)\dot \p_
I \dot \p_J\right]  
\label{Interaction matrix}
\end{eqnarray}
where in the second equality we have substituted the explicit DBI Lagrangian, and used $\dot{H}=-X/c_s$ as well as $c_s^2 = 1-2fX$.

\subsection{Two-field DBI}

We can gain a better intuition for the system of perturbations described by the action (\ref{2d-order-action}) by restricting our attention to a two-field system, $I=1,2$.
Then one can unambiguously decompose perturbations into (instantaneous) 
adiabatic and entropic modes by projecting respectively, parallel and perpendicular to the background trajectory in field space. In other words, we introduce the basis $\{e_\sigma, e_s\}$ where 
$e^I_\sigma=e^I$, and $e^I_s$ is the entropy unit vector orthogonal to $e^I_\sigma$:
\beq
e_\sigma^I\equiv e^I, \qquad G_{IJ}e_s^I e_s^J=1, \qquad G_{IJ}e_s^I e_\sigma^J=0.
\eeq
We also define
\beq
\dot\sigma\equiv \sqrt{2X}.
\eeq
One can reformulate the background equations of motion  Eq.~(\ref{KG}) for DBI in terms of these quantities. The adiabatic component is
\beq
\ddot\sigma=c_s^2\left( c_s \tP_{,\s}-c_s \dot\s^2 \tP_{,\tX \s}-3H\dot\s\right),
\eeq
whereas the entropy component gives the time variation of $e_\s^I$:
\beq
{\mathcal D}_t e^I_\s=\frac{c_s \tP_{,s}}{\dot\sigma}e^I_s\,.
\eeq
In the above equations, $\tP$ is given in (\ref{Ptilde}) and  partial derivatives with respect to $\sigma$ or $s$ denote the projection of the field space gradients along $e_\s^I$ or $e_s^I$ respectively.  For example, $\tP_{,\s} \equiv  e_{\s}^I \tP_{,I}$ and $\tP_{;ss} \equiv e_s^I e_s^J \mathcal{D}_I \mathcal{D}_J \tP$.

On introducing the decomposition 
\beq
\label{decomposition}
Q^I=Q_\s e^I_\s+Q_s e^I_s\,,
\eeq
the equations of motion for $Q_\s$ and $Q_s$ follow from Eqs.~(\ref{2d-order-action}) and (\ref{Interaction matrix}) (see \cite{ds}).  For the adiabatic part one finds
\begin{eqnarray}
\label{Qsigma}
 \ddot{Q}_{\sigma}+\left(3H-3\frac{\dot c_s}{c_s}\right)
 \dot{Q}_{\sigma}+\left(\frac{c_s^2 k^2}{a^2}+\mu_{\s}^2\right)  Q_{\sigma} \, =
\Tdot{\left(\Xi Q_s\right)}
-\left(\frac{\Tdot{(H c_s^2)}}{H c_s^2}
-\frac{c_s \tP_{,\sigma}}{\dot \s }\right)  \Xi\, Q_s\,, 
\end{eqnarray}
where the coupling $\Xi$ between the adiabatic and entropy components is 
\be
\Xi \equiv \frac{c_s}{\dot \s}\left[(1+c_s^2)\tP_{,s} -c_s^2 \tP_{,\tX s}\dot\s^2\right]
 = -c_s\sqrt{\frac{f}{1-c_s^2}}\left[\frac{(1-c_s)^2}{f^2}f_{,s}+(1+c_s^2)V_{,s}\right]
\ee
while the effective mass of the adiabatic modes can be written in the form
\begin{eqnarray}
\mu_{\s}^2 &\equiv & -\frac{\Tddot{(\dot\s/H)}}{\dot\s/H}-\left(3H 
-3\frac{\dot c_s}{c_s}
+\frac{\Tdot{(\dot\s/H)}}{\dot\s/H}\right)
\frac{\Tdot{(\dot\s/H)}}{\dot\s/H}
\label{mu_sigma}\, .
\end{eqnarray}
The equation of motion for the entropy part can be expressed as
\beq
\ddot{Q}_s+\left(3H-\frac{\dot c_s}{c_s}\right)\dot{Q}_s+\left(\frac{ k^2}{a^2}+\mu_s^2+\frac{\Xi^2}{c_s^2}\right) Q_s=-\frac{\dot\s}{\dot H}\Xi \frac{k^2}{a^2} \Psi\,,
\label{delta_s}
\eeq
where the right hand side depends on the Bardeen potential $\Psi$, introduced in Eq.~(\ref{Bardeen}),
and which depends on $Q_\s$ and $Q_s$ through Eqs.~(\ref{supercomplicated}) and (\ref{dN}).  The effective mass appearing above is given by
\begin{eqnarray}
\mu_s^2 &\equiv& -c_s \tP_{;ss}+\half \dot \s^2{{\cal R}}_G-\frac{\tP_{,s}^2}{\dot \s^2}+2 c_s^2 \tP_{,\tX s} \tP_{,s}
\nonumber \\
& =&c_s V_{;ss}-\frac{f}{1-c_s^2}V_{,s}^2-\frac{(1-c_s)^3}{4(1+c_s)f^3} f_{,s}^2-\frac{(2+c_s)(1-c_s)}{(1+c_s)f}f_{,s} V_{,s}
-\frac{(1-c_s)^2}{2f^2} f_{;ss}+\half \dot \s^2{{\cal R}}_G\,.
\label{mus2}
\end{eqnarray}
(${{\cal R}}_G$ is the scalar Riemann curvature in field space.) Note that in this form, Eq.~(\ref{delta_s}) is useful on large scales when the right hand side can be neglected --- in this case one sees immediately that the entropy perturbation $Q_s$ evolves independently of the adiabatic mode.

In order to study the quantum fluctuations of the system, it is convenient, after going to conformal time $\tau = \int {{\rm d}t}/{a(t)}$, to work in terms of canonically normalized fields given by
\be
v_{\s}=\frac{a}{c_s^{3/2}} \, Q_{\s} \,,\qquad \,v_{s}=\frac{a}{\sqrt{c_s}}\, Q_s\,.
\label{v}
\ee
Remarkably, in terms of these new variables, the second order action  (\ref{2d-order-action}) reduces to the very simple
form 
\begin{eqnarray}
\label{S_v}
S_{(2)}&=&\frac{1}{2}\int {\rm d}\tau\,  {\rm d}^3x \Big\{ 
  v_\s^{\prime\, 2}+ v_s^{\prime\, 2} -2\xi v_\s^\prime v_s-c_s^2 \left[(\partial v_\s)^2 + (\partial v_s)^2\right] 
\cr
&& 
\left.
\qquad
+\frac{z''}{z} v_\s^2
+\left(\frac{\alpha''}{\alpha}-a^2 \mu_s^2\right) v_s^2+2\, \frac{z'}{z}\xi v_\s v_s\right\}
\end{eqnarray}
where a prime denotes a derivative with respect to conformal time.  The coupling between $v_\s$ and $v_s$ depends on
\beq
\xi = \frac{a}{c_s }\Xi
\label{11}
\eeq
and we have introduced the two background-dependent  functions 
\beq
z=\frac{a \dot \s }{H c_s^{3/2}}, \qquad \alpha=\frac{a}{\sqrt{c_s}}.
\eeq
This result is similar to that of \cite{ds}, except for the spatial gradient terms which have the same coefficient $c_s^2$ for both the 
adiabatic and isocurvature perturbations. 
The equations of motion for $v_\s$ and $v_s$ are  
\begin{eqnarray}
v_{\s}''-\xi v_{s}'+\left(c_s^2 k^2-\frac{z''}{z}\right) v_{\s} -\frac{(z \xi)'}{z}v_{s}&=&0\,,
\label{eq_v_sigma}
\\
v_{s}''+\xi  v_{\s}'+\left(c_s^2 k^2- \frac{\alpha''}{\alpha}+a^2\mu_s^2\right) v_{s} - \frac{z'}{z} \xi v_{\s}&=&0\,.
\label{eq_v_s}
\end{eqnarray}

In the following we will assume that the time evolution of  $H$, $\dot \s$ and $c_s$ is very slow with respect to that of the scale factor, so that $z''/z\simeq \alpha''/\alpha \simeq 2/\tau^2$. Since $\tau$ varies from $-\infty$ to $0$, 
the wavelength of a given mode is first inside the sound horizon (when $|kc_s\tau| \gg 1$) and then crosses out the sound horizon. 
As in standard inflation, the initial conditions for the perturbations are determined by choosing the familiar Minkowski-like vacuum on very small scales. Below, we consider in turn the quantization on  sub-horizon scales and then the classical evolution on large scales.

\subsubsection{Quantization}
For simplicity, we  assume that  the coupling $\xi$ is very small when the scales of interest cross out the sound horizon, in which case one can quantize the two degrees of freedom independently and solve analytically the system (otherwise, one can resort to numerical integration by starting deep enough inside the sound horizon as in \cite{Tsujikawa:2002qx,Lalak:2007vi}).
The amplification of the vacuum fluctuations at horizon crossing is possible only for very light degrees of freedom. Consequently, if $\mu_s^2$ is larger than $H^2$, this amplification is suppressed and there is no production of entropy modes. 
Interestingly we see from Eq.~(\ref{mus2}) that the term coming from the second derivative of the potential along the entropy direction is multiplied by the sound speed $c_s$, which implies that, with a similar potential, it is easier to generate entropy modes in DBI inflation than in standard slow-roll inflation.   Moreover, the second and third terms in $\mu_s^2$ are always negative and thus tend to destabilize the entropic direction.
Below we assume that $|\mu_s^2|/H^2\ll 1$.

Following the standard procedure (see e.g. \cite{mfb} or \cite{Langlois:2004de})
  one selects the  positive frequency solutions of Eqs.~(\ref{eq_v_sigma}) and (\ref{eq_v_s}), which  correspond to the usual vacuum on very small scales:\beq
v_{\s\, k} \simeq v_{s\, k} \simeq  \frac{1}{\sqrt{2k c_s}}e^{-ik c_s \tau }\left(1-\frac{i}{k c_s\tau}\right)\, .
\eeq
As a consequence, the power spectra for $v_\s$ and $v_s$ after sound horizon crossing  have the same amplitude\beq
{\cal P}_{v_\s}={\cal P}_{v_s}=\frac{k^3}{2\pi^2}|v_{\s\, k}|^2\simeq\frac{H^2 a^2}{4\pi^2 c_s^3}.
\eeq
However, in terms of the initial fields $Q_\s$ and $Q_s$, one finds, using (\ref{v}), 
\beq
\label{power_sigma}
{\cal P}_{Q_\s*}\simeq\frac{H^2}{4\pi^2 }, \quad {\cal P}_{Q_s*}\simeq\frac{H^2}{4\pi^2 c_s^2},
\eeq
(the subscript $*$ indicates that the corresponding quantity is evaluated at sound horizon crossing $k c_s=aH$)
which shows that, for small $c_s$, the entropic modes are {\it amplified} with respect to the adiabatic modes:
\beq
Q_{s*}\simeq \frac{Q_{\sigma*}}{c_s}.
\eeq
In order to confront the predictions of inflationary models to cosmological observations, it is useful to rewrite the scalar field perturbations in terms of geometrical quantities, such as the
 comoving curvature perturbation. The latter is related to the adiabatic perturbation by the expression (\ref{calRdef}), which yields 
 \be
{\cal R}=\frac{H}{\dot \s}Q_{\s}\,,
\label{R}
\ee
so that one recovers the usual {\it single-field} result \cite{Garriga:1999vw} that the power spectrum for $\R$ at sound horizon crossing is given by
\be
{\cal P}_{\cal R_*}=\frac{k^3}{2\pi^2}\frac{|v_{\s\, k}|^2}{z^2}\simeq\frac{H^4}{4\pi^2  \dot{\sigma}^2 }=\frac{H^2}{8\pi^2 \epsilon \, c_s }\,,
\label{power-spectrum-R}
\ee
where $\epsilon \equiv -\dot H / H^2$. 

It is then convenient to define an entropy perturbation, which we denote ${\cal S}$, such that its power spectrum
at sound horizon crossing is the same as that of the curvature perturbation:
\be
{\cal S}=c_s\frac{H}{\dot \s}Q_{s}.
\label{S}
\ee
We thus have
\beq
{\cal P}_{\cal R_*}={\cal P}_{\cal S_*}\equiv {\cal P}_{*}.
\eeq
We stress that our convention for the definition of ${\cal S}$ is purely for convenience.

In contrast with the scalar perturbations, the tensor modes are, as usual, amplified at {\it Hubble radius} crossing.  The amplitude of their power spectrum, given by 
\be
{\cal P}_{\cal T}=\left(\frac{2H^2}{\pi^2}\right)_{k=aH}\,,
\label{power-spectrum-T}
\ee
is much smaller than the curvature amplitude in the small $c_s$ limit.

\subsubsection{Evolution on large scales}
In order to determine the observational consequences of {\it single-field} inflation models, it is usually sufficient to evaluate the amplitude of the comoving curvature perturbation just after horizon crossing. The reason is that the comoving curvature perturbation is conserved on large scales for adiabatic perturbations, as is also the case for the curvature perturbation on uniform energy density hypersurfaces $\zeta$, which coincides with $-\R$ on large scales.  This property is simply a consequence of the conservation of the energy-momentum tensor \cite{Wands:2000dp} (this is also true for non-linear perturbations \cite{Lyth:2004gb,Langlois:2005ii,Langlois:2005qp}).

In contrast with the single-field case, the curvature 
perturbation generally evolves in time, even on large scales, in a multi-field scenario \cite{sy} (see also \cite{Lalak:2007vi} for a recent analysis with non-standard kinetic terms). This can be 
interpreted as due to a transfer between the adiabatic and entropic modes, 
governed by the relation \cite{ds}:
\be
\dot{ \mathcal{R}}=\frac{\Xi}{ c_s} {\cal S} +\frac{H}{\dot H}\frac{c_s^2 k^2}{a^2}\Psi
 \,.
\ee
Note that, whereas this relation might be useful during inflation, it is  not always  relevant, for example at the end of inflation and during reheating where $\dot{\s}$ may temporarily vanish. 
The importance of the transfer depends on the specific  model under consideration and can be computed analytically only in some simple cases.

On large scales the curvature-entropy evolution  can be approximated by two equations of the form 
\be
\dot {\cal R}\approx\alpha H {\cal S}\,,  \qquad \dot {\cal S}\approx\beta H {\cal S}
\label{R-S-evolution},
\ee
where in the latter, we have neglected the second-order time derivative in Eq.~(\ref{delta_s}).  In our case, the coefficients $\alpha$ and $\beta$ are given by 
\be
\alpha= \frac{\Xi}{c_s H} \; ,\qquad \beta \simeq \frac{s}{2}-\frac{\eta}{2}-\frac{1}{3H^2}\left(\mu_s^2+\frac{\Xi^2}{c_s^2}\right),
\label{coefficients}
\ee
where we have introduced the slow-varying parameters
\beq
\eta=\frac{\dot \epsilon}{H \epsilon}\,, \qquad s=\frac{\dot c_s}{H c_s}\,,
\eeq
and kept only the leading order terms in the expression for $\beta$.

The system of equations (\ref{R-S-evolution}) can be formally integrated (see \cite{Wands:2002bn}) to yield
\be
\left( \begin{array}{ccc} 
 {\cal R} \\ 
 {\cal S} 
\end{array} \right)  = 
\left( \begin{array}{ccc} 
1& T_{ {\cal R}  {\cal S} } \\ 
0& T_{ {\cal S}  {\cal S} } 
\end{array} \right) 
\left( \begin{array}{ccc} 
 {\cal R} \\ 
 {\cal S}
\end{array} \right)_{*}
\label{transfer}
\ee
with
\be
T_{ {\cal S}  {\cal S} } (t_{*},t)={\rm exp}\left(\int_{t_{*}}^t \, \beta(t') H(t') {\rm d}t'\right),\qquad T_{ {\cal R}  {\cal S} } (t_{*},t)=\int_{t_{*}}^t \, \alpha(t') T_{ {\cal S}  {\cal S} } (t_{*},t') H(t') {\rm d}t'\,.
\ee
Hence  the (time-dependent) power spectra for the curvature perturbation, the entropy perturbation and the 
correlation between the two can be formally expressed as 
\be
{\cal P}_{\cal R}=(1+T_{{\cal R} {\cal S}}^2) {\cal P}_{{*}}\, , \qquad
{\cal P}_{\cal S}=T_{{\cal S} {\cal S}}^2 {\cal P}_{{*}}\, , \qquad {\cal C}_{{\cal R}  {\cal S}}\equiv \langle {\cal R} {\cal S} \rangle =T_{{\cal R} {\cal S}}   T_{{\cal S} {\cal S}}  {\cal P}_{{*}}\,,
\label{observed-spectrum}
\ee
(recall that ${{\cal R}}$ and ${{\cal S}}$ are implicitly assumed to be uncorrelated at sound horizon crossing).

An interesting question, which depends on the details of reheating and thus goes beyond the scope of the present work, is whether the entropy perturbation {\it during} inflation can be transfered to some entropy perturbations {\it after} inflation, i.e.~in the radiation phase. If this is the case, then the primordial entropy fluctuations could be directly observable, with the interesting possibility that there could be a correlation between the adiabatic and entropy modes 
\cite{Langlois:1999dw}. 

In any case, one can introduce the correlation angle $\Theta$, defined by
\be
{\sin} \Theta
\equiv \frac{{\cal C}_{{\cal R}  {\cal S}}}{\sqrt{{\cal P}_{\cal R}}\sqrt{{\cal P}_{\cal S}} }
\label{correlation angle}
\ee
which can also be seen as a transfer angle, since  
\be
{\sin} \Theta =\frac{T_{ {\cal R}  {\cal S} }}{\sqrt{1+T^2_{ {\cal R}  {\cal S} }}}\,.
\label{correlation-result}
\ee
If $\Theta=0$ there is no transfer ($T_{ {\cal R}  {\cal S}} =0$), whereas if $|\Theta|=\pi/2$ ($T_{ {\cal R}  {\cal S} }\gg 1$) the final curvature perturbation is mostly of entropic origin.
The relationship between the curvature power spectrum at sound horizon crossing and its final value is thus
\be
{\cal P}_{\cal R_{*}}={\cal P}_{\cal R} {\cos^2} \Theta.
\label{PR-initial}
\ee
This implies, on using the tensor amplitude Eq.~(\ref{power-spectrum-T}), that  the tensor to scalar ratio is given by
\be
r \equiv \frac{{\cal P}_{\cal T}}{{\cal P}_{\cal R}}=16 \, \epsilon \, c_s \,{\cos^2} \Theta.
\label{r}
\ee
Interestingly this expression combines the result of $k$-inflation \cite{Garriga:1999vw}, where the ratio is suppressed by the sound speed $c_s$, and that of standard multi-field inflation \cite{Wands:2002bn}.

From the expression of the curvature power spectrum,  
one can compute the scalar spectral index  in the  slow-varying approximation.  We obtain
\be
n_{\cal R}\equiv \frac{{\rm d\,ln}{\cal P_{{\cal R}}}}{{\rm d\, ln\,}k}=n_{\cal R_*}+H_*^{-1} {\rm sin}(2 \Theta) \frac{\partial T_{ {\cal R}  {\cal S} }}{\partial t^*} =n_{\cal R_*}-\alpha_{*}{\rm sin}(2 \Theta)-2\beta_*{\rm sin^2} \Theta
\label{indices}
\, 
\ee
with
\be
n_{\cal R_*}-1=-2\epsilon_*- \eta_* -s_* \,,
\ee
and where we have used %
\be
H_*^{-1}\frac{\partial T_{ {\cal S}  {\cal S}}}{\partial t^*}=-T_{{\cal S}  {\cal S} } \beta_*\,, \qquad H_*^{-1}\frac{\partial T_{ {\cal R}  {\cal S}}}{\partial t^*}=-\alpha_*-T_{{\cal R}  {\cal S} } \beta_* \,.
\label{derivative-transfer}
\ee
The observable spectral index, given in Eq.~(\ref{indices}), not only depends on the values of the various parameters at sound horizon crossing, but also  on  the transfer angle $\Theta$.

\section{Non-Gaussianities}
\label{sec:6}

In the simplest models of inflation, primordial perturbations are characterized by a very small amount of non-Gaussianity \cite{Maldacena:2002vr}. However, other models, such as single-field DBI inflation, are expected to produce significant non-Gaussianity \cite{Bartolo:2004if}. If ever detected, primordial non-Gaussianity would be a powerful discriminator between various 
early universe models. 
In order to study non-Gaussianities, one must analyse the perturbations beyond linear order. During inflation, primordial non-Gaussianities can arise from the quantum fluctuations at horizon crossing or, in the case of multi-field inflation,  from the classical non-linear evolution on large scales (see e.g.~\cite{Lyth:2005fi,Vernizzi:2006ve}).

In this section, we concentrate on the primordial non-Gaussianity originating from the three-point function 
of the scalar field fluctuations, which is the main contribution in
single-field DBI inflation. Its calculation  requires the third-order action in  perturbations. Below, we first consider the general case --- that is  models of the form (\ref{P_XIJ}) --- and then specialise to DBI.

\subsection{Third-order action: general case}

We follow the standard approach which has been presented in \cite{Maldacena:2002vr,Seery:2005wm,Seery:2005gb,Chen:2006nt}, considering successively the third-order action from the Einstein-Hilbert term and then from the matter part. A similar calculation of the third-order action can be found in \cite{Gao:2008dt}, but only for the multi-field
Lagrangians of the form $P(X,\phi^K)$, where $X=\delta_{IJ}X^{IJ}$.

The third-order action coming from the gravitational part is the same as in the single field case and is given by 
the expression
\be
S_{(3)}^{(G)} = \frac{1}{2} \int  {\rm d}t \, {\rm d}^3 x \, a^3 \left\{-\frac{\delta N}{a^4}\left[\left(\partial_i\partial_j\pp \right)
\partial^i\partial^j\pp-\left(\partial^2\pp\right)^2\right]+4\frac{H}{a^2}\partial^2\pp(\delta N)^2 +6H^2(\delta N)^3  \right\}\,,
\ee
where the relation between $\delta N$ and the field perturbations given in Eq.~(\ref{dN}) can be rewritten as  
\be
\delta N \equiv \N_A Q^A
\ee
with the (field-space) vector 
\be
\label{NA}
\N_A = \frac{1}{2H} P_{<AB>}\dot{\phi}^B.
\ee
By expanding systematically  the matter part of the action up to third order, we finally find (intermediate steps can be found in Appendix 3)
\ba
S_{(3)}^{(M)} &=& \int {\rm d}t \, {\rm d}^3 x\,  a^3 \left( \delta^{(3)} P + (\delta N) \delta^{(2)}P \right)
\nn
\\
&=& \int  {\rm d}t\,  {\rm d}^3 x \, a^3 \left\{ (g_1)_{ABC} Q^A Q^B Q^C  +  (g_2)_{ABC}Q^A Q^B \dot{Q}^C +
(g_3)_{ABC}Q^A \dot{Q}^B \dot{Q}^C
\right.
\nn
\\
&&+ 
\left. (g_4)_{ABC}\dot{Q}^A \dot{Q}^B \dot{Q}^C +(g_a)_{AB}Q^A \partial_jQ^B (\delta N^j)  + (g_b)_{AB} \dot{Q}^A \partial_jQ^B (\delta N^j)  
\right.
\nn
\\
&&+ 
\left. (g_c)_{ABC} Q^A (h^{ij} \partial_i{Q}^B \partial_j{Q}^C)  + (g_d)_{ABC}\dot{Q}^A (h^{ij} \partial_i{Q}^B \partial_j{Q}^C) 
 \right\}
\ea
with
\ba
 (g_1)_{ABC}  &=& \frac{1}{6} P_{,ABC} + \frac{1}{2}P_{,BC} \N_A -  P_{<IJ>}X^{IJ}\N_A \N_B \N_C
 \nn
 \\
  && + P_{<IJ>,B} X^{IJ} \N_A \N_C  - 4   P_{<IJ>,<KL>}X^{IJ} X^{KL} \N_A \N_B \N_C - X^{IJ}P_{<IJ>,BC}\N_A
\nn
\\
&& +  2P_{<IJ>,<KL>,C}X^{IJ} X^{KL} \N_A \N_B
\nn
\\
&&- \frac{4}{3}  P_{<IJ>,<KL>,<MN>}X^{IJ} X^{KL}X^{MN} \N_A \N_B \N_C
\label{g1}
\ea
\ba
 (g_2)_{ABC}  &=& 2H \N_A \N_B \N_C + \frac{1}{2} P_{<KC>,AB} \dot{\phi}^K
 \nn
 \\
 && + \N_A \dot{\phi}^K \left[ 5 \N_B X^{IJ} P_{<IJ>,<KC>}
- P_{<KC>,B} 
+ 2 \N_B   X^{IJ} X^{NL}P_{<IJ>,<NL>,<KC>}
\right.
\nn
\\
&& 
\left. - 2 X^{IJ}  P_{<IJ>,<KC>,B} \right] 
\label{g2}
\\
&&
\nn
 \\
 (g_3)_{ABC}  &=& -\frac{1}{2} \N_A P_{<BC>} + \frac{1}{2} P_{<BC>,A} - \N_A \left[
 3X^{IK} P_{<IB>,<KC>} + X^{KL} P_{<BC>,<KL>} \right]
 \nn
 \\
 && - 2 \N_A  P_{<IJ>,<KB>,<MC>}X^{IJ} X^{KM} + P_{<IB>,<KC>,A}
X^{IK} 
\label{g3}
\\
&&
\nn
\\
 (g_4)_{ABC}  &=&
\frac{1}{2} \dot{\phi}^M P_{<BC>,<MA>} +\frac{1}{3} X^{IK} \dot{\phi}^M P_{<IA>,<KB>,<MC>} 
\label{g4}
\\
&&
\nn
\\
 (g_a)_{AB}  &=& 2H \N_A \N_B + 2 \N_A X^{KL}\dot{\phi}^I P_{<IB>,<KL>}  - \dot{\phi}^I P_{<IB>,A}
 \label{ga}
 \\
&&
\nn
\\
 (g_b)_{AB}  &=& -P_{<AB>}  - 2X^{IK} P_{<IB>,<KA>}
 \label{gb}
 \\
&&
\nn
\\
 (g_c)_{ABC} & = & \N_A X^{KL}P_{<BC>,<KL>} - \frac{1}{2} P_{<BC>,A} - \frac{1}{2} P_{<BC>}\N_A
 \label{gc}
 \\
&&
\nn
\\
 (g_d)_{ABC}  &=& - \frac{1}{2} \dot{\phi}^K P_{<BC>,<KA>} \,.
 \label{gd}
 \ea

In order to get a flavour for the new effects which could arise in multi-field inflation with {\it non-standard} kinetic terms, it is instructive to compare the above terms with their counterparts in standard multi-field inflation, such as studied in \cite{Seery:2005gb}. Substituting  the standard matter Lagrangian 
\beq
P= G_{IJ}X^{IJ}-V(\phi),\qquad  G_{IJ}=\delta_{IJ} \, ,
\eeq
the above coefficients reduce to 
\ba
&& (g_1)_{ABC}  = -\frac{1}{6} V_{,ABC} - \frac{1}{2}V_{,BC} \N_A - X\N_A \N_B \N_C, \cr
&& (g_2)_{ABC}  = 2H \N_A \N_B \N_C, \quad
 (g_3)_{ABC}  = -\frac{1}{2} \N_A G_{BC}, \quad 
 (g_4)_{ABC}  =0, 
\cr
&& (g_a)_{AB}  = 2H \N_A \N_B, \quad
 (g_b)_{AB}  = -G_{AB}, \quad
 (g_c)_{ABC}  =  - \frac{1}{2} G_{BC}\N_A, \quad
 (g_d)_{ABC}  = 0.
 \ea 
As we will see in the next subsection, the main contribution for DBI inflation will come precisely from the vertices associated 
with the coefficients $g_4$ and $g_d$, which do not exist for standard kinetic terms.

\subsection{Non-Gaussianities in DBI inflation}

Single-field DBI inflation is an inflationary model which naturally produces a (relatively) high level of non-Gaussianity in the small $c_s$ limit, as shown in \cite{ast04,Chen:2005fe}. 
It is thus important to investigate how the amplitude and shape of primordial 
non-Gaussianities are modified in the multi-field case \cite{Babich:2004gb,Bernardeau:2002jy}. 

Here, we will focus on the dominant contributions to the non-Gaussianities for $c_s \ll 1$, and therefore ignore the contributions coming from the gravitational part of the action, which are known to be sub-dominant.
As in the single-field case, the dominant contributions
come from the terms  involving derivatives of $P$ with respect to the $X^{IJ}$'s, because they are enhanced  by negative powers of $c_s$ with respect to the other terms. Moreover, terms containing ${\cal N}_A$ are suppressed in the slow-varying approximation. Indeed, if one compares for example the first term of $g_c$ with $g_d$, one finds schematically
\beq
\frac{g_c}{H g_{d}}\sim \frac{\dot\sigma^2}{H^2 c_s},
\eeq
which is proportional to $\epsilon=-\dot H/H^2$ and thus small. 
Finally, the dominant contributions come from the following terms in $S_{(3)}$: 
\beq
(g_4)_{IJK}\dot Q^I\dot Q^J\dot Q^K+ (g_d)_{IJK}\, \dot Q^I h^{jk} \partial_j Q^J \partial_k Q^K\, .
\eeq
On substituting the multi-field DBI Lagrangian into the expressions given in Eqs.~(\ref{g4}) and (\ref{gd}), the coefficients 
$g_4$ and $g_d$ can be calculated explicitly (one needs the third derivative of $P$ with respect to the
$X^{IJ}$, which can be deduced from  Appendix 1).

In the two-field case, decomposing the fields in terms of their adiabatic and entropic components according to 
(\ref{decomposition}), as well as using (\ref{P(IJ)(KL)}), one finally finds that the relevant terms of the third-order action are given by
\ba
S_{(3)}^{(\rm main)}&=&\int {\rm d}t\, {\rm d}^3x\,  \left\{ \frac{a^3}{2 c_s^5 \dot \s}\left[(\dot Q_{\s} )^3+c_s^2 \dot Q_{\s}  (\dot Q_{s} )^2\right]
 - \frac{a}{2 c_s^3 \dot \s}\left[ \dot Q_{\s} (\nabla  Q_{\s} )^2
 -c_s^2 \dot{Q_{\s} }(\nabla Q_s)^2+2 c_s^2 \dot {Q_s}\nabla Q_{\s} \nabla Q_s)\right]
 \right\}
 \label{S3}
\ea
where we have used the fact that $f\simeq 1/\dot\sigma^2$ in the limit $c_s\ll 1$.
All the terms which appear in Eq.~(\ref{S3}) are of the same order of magnitude,  since $Q_s \simeq Q_{\s}/c_s$ as we have seen earlier. 
Note that using $X$ instead of  $\tilde{X}$  in the DBI action (that is, neglecting the higher order terms appearing in $\H$) would lead to a different third-order action.

Let us now compute the contribution of these vertices to the relevant three-point functions, by following the procedure outlined in detail in  \cite{Seery:2005wm}. Working at leading order in the slow varying regime,  we use the adiabatic and entropic propagators defined by, respectively, 
\be
\langle Q_{\s}(0)Q_{\s}(\tau)\rangle=\frac{H^2}{2k^3}(1-i k c_s \tau){\rm e}^{i k c_s \tau} \, , \qquad   \langle Q_{s}(0)Q_{s}(\tau)\rangle=\frac{H^2}{2k^3 c_s^2}(1-i k c_s \tau){\rm e}^{i k c_s \tau} \,,
\ee
which correspond to the Fourier transforms of the Green functions, solutions of Eqs.~(\ref{eq_v_sigma}) and (\ref{eq_v_s}) with $\xi=0$ 
and $z''/z=\alpha''/\alpha = 2/\tau^2$.
The calculation of the three-point functions involve time integrations and we assume that, as usual, the main contribution to these integrals comes from the period around horizon crossing \cite{Weinberg:2005vy},  which enables us to extrapolate the integration bound to $\tau=0$. We  also implicitly ignore the correlations at  different times between the adiabatic and entropy modes, since these are expected to be small if the coupling $\xi$ is small. The quantities 
$\dot \s$ and $c_s$ will be considered as constant in time in the integrals. 
Given these assumptions, the only integrals required are 
\be
\int_{-\infty}^0 {\rm d} \tau {\rm e}^{i K c_s \tau}=-\frac{i}{K c_s}, \qquad \int_{-\infty}^0 {\rm d} \tau\,  \tau {\rm e}^{i K c_s \tau}=\frac{1}{(K c_s)^2}, \qquad \int_{-\infty}^0 {\rm d} \tau\, \tau^2 {\rm e}^{i K c_s \tau}=\frac{2i}{(K c_s)^3} 
\label{integrals}
\ee
which have be computed by using the appropriate contour in the complex plane ($\tau \to -(\infty- i \epsilon)$).

The contributions to the three-point function $\langle Q_{\s}  (\boldsymbol{k}_1) Q_{\s}  (\boldsymbol{k}_2) Q_{\s}  (\boldsymbol{k}_3)\rangle$ are, respectively, 
\be
(2 \pi)^3 \delta (\sum_i \bk_i) \frac{3H^4}{2\sqrt{2 c_s \epsilon} c_s^2}\frac{1}{\prod_i k_i^3}  \frac{k_1^2 k_2^2 k_3^2}{K^3}
\label{vertex-1}
\ee
from the vertex proportional to $\dot{Q}_{\s}^3$ and 
\be
-(2 \pi)^3 \delta (\sum_i \bk_i) \frac{H^4}{4 \sqrt{2 c_s \epsilon} c_s^2}\frac{1}{\prod_i k_i^3} \left[ k_1^2( \bk_2 \cdot \bk_3) \left(\frac{1}{K}+\frac{k_2+k_3}{K^2}+\frac{2 k_2 k_3}{K^3}\right) + {\rm perm.} \right]
\label{vertex-3}
\ee
from the vertex proportional to $\dot{Q_{\s} }(\nabla Q_{\s} )^2$, where we have introduced $K\equiv k_1+k_2+k_3$ and used $\dot\sigma=H\sqrt{2\epsilon c_s}$. Summing these contributions, one thus finds 
\ba
\langle Q_{\s}  (\boldsymbol{k}_1) Q_{\s}  (\boldsymbol{k}_2) Q_{\s}  (\boldsymbol{k}_3)\rangle
&=&-(2 \pi)^3 \delta (\sum_i \bk_i) \frac{H^4}{4\sqrt{2 c_s \epsilon} c_s^2}\frac{1}{\prod_i k_i^3 K^3}  \left[ -6 k_1^2 k_2^2 k_3^2
\right.
\cr
&+&\left.
 k_3^2 (\bk_1 \cdot \bk_2)(2 k_1 k_2 -k_3 K +2 K^2)  + {\rm perm.}\right]
\label{usual}
\ea
 where the `perm.' indicate two other terms with the same structure as the last term but permutations of indices 1, 2 and 3).
This is the standard result from single field DBI inflation \cite{Chen:2005fe}.

Let us now turn to the new terms which arise from the entropy fluctuations. They appear in the 
three-point function  $\langle Q_{\s}  (\boldsymbol{k}_1) Q_s (\boldsymbol{k}_2) Q_s (\boldsymbol{k}_3)\rangle$, with the contribution
\be
(2 \pi)^3 \delta (\sum_i \bk_i) \frac{H^4}{2\sqrt{2 c_s \epsilon} c_s^4}\frac{1}{\prod_i k_i^3}  \frac{k_1^2 k_2^2 k_3^2}{K^3}
\label{vertex-2}
\ee
from the vertex proportional to $\dot{Q}_{\s} \dot{Q}_s^2$,
the contribution 
\be
(2 \pi)^3 \delta (\sum_i \bk_i) \frac{H^4}{4 \sqrt{2 c_s \epsilon} c_s^4}\frac{1}{\prod_i k_i^3} k_1^2( \bk_2 \cdot \bk_3) \left(\frac{1}{K}+\frac{k_2+k_3}{K^2}+\frac{2 k_2 k_3}{K^3}\right)
\label{vertex-4}
\ee
from the vertex proportional to $\dot{Q_{\s} }(\nabla Q_s)^2$
and finally the contribution 
\be
-(2 \pi)^3 \delta (\sum_i \bk_i) \frac{H^4}{4 \sqrt{2 c_s \epsilon} c_s^4}\frac{1}{\prod_i k_i^3} \left[k_3^2( \bk_1 \cdot \bk_2) \left(\frac{1}{K}+\frac{k_1+k_2}{K^2}+\frac{2 k_1 k_2}{K^3}\right) + (k_2 \leftrightarrow k_3)\right]
\label{vertex-5}
\ee
from the vertex proportional to $\dot{Q_s}\nabla Q_s\nabla Q_{\s} $.

Summing these three contributions, we find 
\begin{eqnarray}
\langle Q_{\s}  (\boldsymbol{k}_1) Q_s (\boldsymbol{k}_2) Q_s (\boldsymbol{k}_3)\rangle  
&=& -(2 \pi)^3 \delta (\sum_i \bk_i) \frac{H^4}{4\sqrt{2 c_s \epsilon} c_s^4}\frac{1}{\prod_i k_i^3 K^3} 
\left[ -2 k_1^2 k_2^2 k_3^2
\right.
\nn
\\
&&
- k_1^2 (\bk_2 \cdot \bk_3)(2 k_2 k_3 -k_1 K +2 K^2)
\nn
\\
&&
+k_3^2 (\bk_1 \cdot \bk_2)(2 k_1 k_2 -k_3 K +2 K^2) 
\nn
\\
&&\left.
+k_2^2 (\bk_1 \cdot \bk_3)(2 k_1 k_3 -k_2 K +2 K^2) \right].
\end{eqnarray}
As we will see below,  the three-point function of the curvature perturbation depends on the symmetrized (with respect to  permutations of the three wave vectors ${\bk}_i$)  version of this three-point function, and this  has 
 exactly the same shape as (\ref{usual}). Nevertheless, its amplitude is enhanced with respect to the purely adiabiatic one by a factor of
$1/c_s^2$.

Let us now relate the correlation function of the scalar fields to the three-point function of the curvature perturbation $\R$ which is the observable quantity.  In order to do so, we use Eqs.~(\ref{R}), (\ref{S}) and (\ref{transfer}) to write
\be
{\cal R} \approx \A_\sigma  Q_{\s*} + \A_s  Q_{s*} 
\ee
with
\be
\A_\sigma = \left( \frac{H}{\dot \s}\right)_*  \qquad
\A_s =  T_{{\cal R} {\cal S}} \left( \frac{ c_s H}{\dot \s}\right)_*\,.
\label{comparision-transfer}
\ee
Let us compute the three-point funtion  for three wave vectors of comparable magnitude (so that the coefficients
$A_{\s}$ and $A_{s}$, which depend on the time at which the relevant scales cross the sound horizon,  have approximately the same value). The two three-point functions of the fields we have calculated give the following contribution 
\begin{eqnarray}
\langle\R (\boldsymbol{k}_1)  \R (\boldsymbol{k}_2)  \R (\boldsymbol{k}_3)\rangle^{(3)} &=& (\A_{\s})^3\langle Q_{\s}  (\boldsymbol{k}_1) Q_{\s}  (\boldsymbol{k}_2)  Q_{\s}  (\boldsymbol{k}_3) \rangle
 \nn
 \\
  && 
+\A_{\s}(\A_{s})^2(\langle Q_{\s}  (\boldsymbol{k}_1)  Q_s (\boldsymbol{k}_2)  Q_s (\boldsymbol{k}_3)\rangle+  {\rm perm.})
\nn
\\ 
&=& 
(\A_{\s})^3\langle Q_{\s}  (\boldsymbol{k}_1) Q_{\s}  (\boldsymbol{k}_2)  Q_{\s} (\boldsymbol{k}_3) \rangle
\left(1+T_{{\cal R} {\cal S}}^2\right)
\label{zeta-3}
\end{eqnarray}
where the adiabatic three-point function is given in Eq.~(\ref{usual}).
Note that the enhancement of the mixed correlation $\langle Q_{\s}  Q_{s} Q_s \rangle$ by a factor of $1/c_s^2$ is compensated by the ratio between $\A_{\s}$ and $\A_{s}$ so that the purely adiabatic and mixed  contributions in (\ref{zeta-3}) are exactly of the same order.

The superscript (3) in the above equation indicates that we take into account only the contribution from the three-point function of the scalar fields.  One could also include the contribution from the four-point function of the scalar fields, which can be expressed in terms of the power spectra using Wick's theorem, and also from other higher-order terms. This has been done for instance in \cite{Lyth:2005fi,Vernizzi:2006ve}. In the single-field DBI case, the corresponding contibution $f_{NL}^{(4)}$ is negligible compared to $f_{NL}^{(3)}$. 
Because of the transfer between adiabatic and entropic modes, this should be reconsidered in specific multi-field models.
Here we simply disregard these contributions though it should be borne in mind that they are present in principle.

Instead of the three-point function, it is now customary to use the non-Gaussianity parameter $f_{NL}$ defined by 
\be
\langle\R (\boldsymbol{k}_1)  \R (\boldsymbol{k}_2)  \R (\boldsymbol{k}_3)\rangle=-(2 \pi)^7 \delta (\sum_i \bk_i) \left(\frac{3}{10}f_{NL}({\cal P}_{\cal R})^2 \right)\frac{\sum_i k_i^3}{\prod_i k_i^3}\,.
\label{def f_NL}
\ee
From the relation between ${\cal P}_{\cal R}$ and ${\cal P}_{\cal R*}$ given in Eq.~(\ref{PR-initial}), we then obtain, for the equilateral configuration,
\be
f_{NL}^{(3)}=-\frac{35}{108}\frac{1}{c_s^2}\frac{1}{1+T^2_{{\cal R} {\cal S}} }=-\frac{35}{108}\frac{1}{c_s^2} {\cos^2} \Theta \,.
\label{f_NL3}
\ee
One can easily understand this result.  The curvature power spectrum is amplified by a factor of $(1+T^2_{{\cal R} {\cal S}})$ due to the feeding of curvature by entropy modes.
Similarly the three-point correlation function for ${\cal R}$ resulting from the three-point correlation functions of the adiabatic and entropy modes is enhanced by the same factor $(1+T^2_{{\cal R} {\cal S}})$. However, since $f_{NL}$ is roughly the ratio of the three-point function with respect to the {\it square} of the power spectrum, one sees that $f_{NL}$ is now {\it reduced} by the factor 
$(1+T^2_{{\cal R} {\cal S}})$.  This may be important in confronting DBI models to observations \cite{Bean:2007hc,Peiris:2007gz}.

We end by revisiting the consistency condition relating  the non-Gaussianity of the curvature perturbation, the tensor to scalar ratio $r$, and the tensor spectral index $n_{\cal T} =-2\epsilon$, given in \cite{Lidsey:2006ia} for single-field DBI.  In our case, substituting $f_{NL}^{(3)} \simeq -\frac{1}{3}\frac{1}{c_s^2} {\cos^2} \Theta$ in (\ref{r}), gives
\be
r+8 n_{\cal T}=-r\left( \sqrt{-3 f_{NL}^{(3)}}\cos^{-3} \Theta-1\right),
\label{consistency}
\ee
As we can can see from (\ref{f_NL3}) and (\ref{consistency}), violation of the standard inflation consistency relation (corresponding to a vanishing right-hand side in (\ref{consistency})) would be stronger in multi-field DBI than in single-field DBI, and thus easier to detect.  In the multi-field case the consistency condition is only an inequality, unless the entropy modes survive after inflation in which case $\Theta$ is potentially observable.

\section{Conclusions}
\label{sec:7}

In this paper we have studied cosmological perturbations in multi-field
inflation models for which the Lagrangian depends {\it a priori} on all the $N(N+1)/2$ kinetic terms that can be constructed 
by contracting the spacetime gradients of the $N$ scalar fields.   
Our analysis can be seen as the multi-field extension of $k$-inflation, and it also generalizes very recent papers which considered more restrictive Lagrangians of the form $P=P(X,\phi^K)$.
In our very general framework, we have computed the second-order action which governs the dynamics of the linear perturbations, and were thus able to identify the propagation matrix whose eigenvalues correspond to the generalized propagation speeds.

We have argued that such a general framework is necessary in order to study multi-field DBI inflation.  In that model, we showed 
that all modes propagate with the same speed of sound $c_s$, and hence (if light) they are all amplified simultaneously at  sound horizon crossing. However, because their respective canonically normalized functions differ, the result is that the entropy modes are {\it enhanced} with respect to the adiabatic modes: $Q_s\sim Q_\s/c_s$. If there is a subsequent transfer from the entropy modes into the curvature perturbation --- a generic feature as soon as the trajectory in field space is non trivial --- the final amplitude of the curvature perturbation is significantly affected by the entropy modes.  

We have also derived, in the general case, the third-order action from which one can compute the predictions for primordial non-Gaussianities. We have identified the vertices which appear in this action and expressed their coefficients in terms of the initial Lagrangian and its derivatives. 
In the  DBI case, we have computed the dominant contributions to the non-Gaussianities of the curvature perturbation in the small $c_s$ limit. If there is an entropy-curvature transfer, we have shown that the contribution from the entropy modes  will increase the amplitude of the three-point function with respect to the single field DBI prediction, but the shape of the non-Gaussianities remains exactly the same. Since the entropy modes enhance the curvature two-point and three-point functions  by the same amount, it implies that the $f_{NL}$ parameter, which is related to the three-point function divided by the square of the two-point function, is smaller than in the single field case. The impact of the entropy modes can be expressed simply in terms of the entropy-curvature transfer coefficient, which is model-dependent.  In the future, it would be interesting  to study specific scenarios of DBI inflation and to estimate quantitatively this transfer coefficient.

\section*{Acknowledgements}

We would like to thank E. Babichev, P. Creminelli, G. Tasinato, K. Turzynski, F. Vernizzi for interesting discussions related to the topic of this paper. We would also like to thank the participants of the workshop ``Cosmological frontiers in fundamental physics'', and especially G. Shiu and E. Silverstein, for their useful questions and comments. 
We also acknowledge the support from a joint CNRS-JSPS grant.  TT is supported by Monbukagakusho Grant-in-Aid for Scientific Research Nos.~17340075 and 19540285.

\section*{Appendix 1: the DBI Lagrangian and its derivatives}
\label{App:1}

\subsection*{Calculation of the DBI determinant}

The expression for ${\cal D}$ in Eq.~(\ref{def_explicit}) can be obtained from its definition in Eq.~(\ref{Ddef}) on substituting into
\beq
{\rm det}(\M)=-\frac{1}{4!}\epsilon_{\a_1\a_2\a_3\a_4}\epsilon^{\b_1\b_2\b_3\b_4}\M^{\a_1}_{\ \beta_1}\M^{\a_2}_{\ \beta_2}\M^{\a_3}_{\ \beta_3}\M^{\a_4}_{\ \beta_4}
\eeq
the matrix of components
\beq
\M^\alpha_\beta=\delta^\alpha_\beta+ f B_I^\alpha 
B_\beta^I, \quad B_I^\alpha\equiv G_{IJ}\partial^\alpha \phi^J, \quad B^I_\beta\equiv \partial_\beta\phi^I.
\eeq
On using the identity
\beq
\epsilon_{\a_1\a_2\a_3\a_4}\epsilon^{\alpha_1\dots\alpha_j\b_{j+1}\dots\b_4}=-(4-j)!\, j!\, \delta^{[\beta_{j+1}}_{\, \a_{j+1}}
\dots \delta^{\beta_4]}_{\,\a_4},
\eeq
and the contractions $B_I^\alpha 
B_\alpha^J=-2 G_{IK}X^{KJ}$, 
one finally gets the expression 
\beq
\label{D_explicit2}
{\cal D}=1-2f G_{IJ}X^{IJ}+4f^2 X^{[I}_IX_J^{J]} -8f^3 X^{[I}_IX_J^{J} X_K^{K]}+16f^4 X^{[I}_IX_J^{J} X_K^{K}X_L^{L]} \, .
\eeq
If there are three scalar fields, the last term disappears because of the antisymmetrization over the field indices. For two scalar fields, the last two terms disappear; and for one scalar field, only the first two terms remain.
For more than four scalar fields, the truncation at order $f^4$ is natural if one considers ${\cal D}$ as the 
determinant of a $4\times 4$ matrix, but it is less obvious if one starts from the expression for  ${\cal D}$ as the determinant of a $N\times  N$ matrix, Eq.~(\ref{det}).  However, this can be understood  by noting that the term proportional to $f^n$ is a sum of products involving  $n$ terms of the form  $X^I_J=-B^I_\mu B_J^\mu/2$. If $n>4$, among 
the $n$ terms of the form $B^I_\mu$, at least two have the same index $\mu$
because the spacetime index $\mu$ can only take four different values. Since, by
definition of the determinant,  all the field indices $I$ are
antisymmetrized, one thus finds that the term of order $f^n$ necessarily vanishes.

\subsection*{Derivatives}
In order to compute the derivatives of the DBI Lagrangian with respect to $X^{IJ}$, one can use the explicit 
expression for ${\cal D}$ given above in Eq.~(\ref{D_explicit2}). An alternative derivation, 
which is simpler, is to start from the identity
\beq
\label{D_pert}
{\cal D}= \exp\left[{\rm Tr}\, {\ln} M\right]=\exp\left[{\rm Tr} \, {\ln} (\bar{M}+\delta M)\right]
=\exp\left\{{\rm Tr}\left[ {\ln} ({\bar M})+{\rm ln}(\mathbf{Id}+\bar{M}^{-1}\delta M)\right]\right\},
\eeq
where the components of $M$, given in (\ref{det2}), are decomposed into
\beq
 \bar{M}_{I}^{\, J}=\delta_{I}^{\, J}-2f G_{IK}\bar{X}^{KJ}\, ,\qquad \delta M_I^J=-2fG_{IK} \delta X^{KJ}.
\eeq
Moreover the components of the matrix $(\bar{M})^{-1}$ are the $\tilde{G}_I^{\ J}$ given in Eq.~(\ref{Gtilde}).

Using (\ref{D_pert}), the expansion of ${\cal D}^{1/2}$ in terms of the matrix $U=\bar{M}^{-1}\delta M$ yields
\ba
{\cal D}^{1/2}&=&\bar{{\cal D}}^{1/2}\exp\left[\frac{1}{2}{\rm Tr}(U)-\frac{1}{4}{\rm Tr}(U^2)+\frac{1}{6}{\rm Tr}(U^3)+\dots  \right]\cr
&=&\bar{{\cal D}}^{1/2}\left[1+\frac{1}{2}{\rm Tr}(U)-\frac{1}{4}{\rm Tr}(U^2)+\frac{1}{8}({\rm Tr}U)^2
+\frac{1}{6}{\rm Tr}(U^3)-\frac{1}{8}{\rm Tr}(U){\rm Tr}(U^2)+\frac{1}{48}({\rm Tr}U)^3+\dots\right].
\ea
Substituting in the above expression the components
\beq
U_I^{\ J}=-2 f\tG_{IK}\delta X^{KJ}
\eeq
of the matrix $U$, one gets
\ba
{\cal D}^{1/2}&=&\bar{{\cal D}}^{1/2}\left[1-f\tG_{IJ}\delta X^{IJ}-\frac{f^2}{2} \left(2 \tG_{IL}\tG_{JK}
- \tG_{IJ}\tG_{KL} \right)\delta X^{IJ}\delta X^{KL}
\right.
\cr
&&
\left.
+f^3 \left(\tG_{IJ}\tG_{KM}\tG_{LN}-\frac{4}{3}\tG_{IN}\tG_{JK}\tG_{LM}-\frac{1}{6}\tG_{IJ}\tG_{KL}\tG_{MN}
\right)\delta X^{IJ}\delta X^{KL}\delta X^{MN}
\right].
\ea
By interpreting this relation as a Taylor expansion with respect to the variables $X^{IJ}$, one can obtain the derivatives of ${\cal D}^{1/2}$ and thus of the DBI Lagrangian with respect to the $X^{IJ}$, up to third 
order, as required for the computation of the non-Gaussianities.

\section*{Appendix 2: variations of $\tX$ up to second order}
\label{App:2}

The computation of the first and second order variations 
of $\tX$ follows from
\beq
\delta^{(1)}\tX=\delta^{(1)} X+\H_{<IJ>}\delta^{(1)}X^{IJ}+\H_{,K}Q^K
\label{delta1}
\eeq
where
\beq
\delta^{(1)}X^{IJ}=\dot{\phi}^{(I}\dot{Q}^{J)}- \dot{\phi}^I\dot\phi^J \delta N\, ,
\label{delta1XIJ}
\eeq
and also from
\ba
\delta^{(2)}\tX&=&\delta^{(2)} X+\H_{<IJ>}\delta^{(2)}X^{IJ}+\frac{1}{2}\H_{<IJ>,<KL>}\delta^{(1)}X^{IJ}\delta^{(1)}X^{KL}
+\frac{1}{2}\H_{,KL}Q^KQ^L+
\H_{<IJ>,K}\delta^{(1)}X^{IJ}Q^K\,
\label{delta2}
\ea
where
\beq
\delta^{(2)}X^{IJ}=\frac{1}{2}\dot Q^I\dot Q^J+\frac{3}{2}\dot{\phi}^I\dot\phi^J\delta N^2+\dot{\phi}^{(I}\delta^{(2)}v^{J)}-\frac{1}{2} h^{ij} \partial_i Q^I\partial_j Q^J-2 \dot{\phi}^{(I}\dot{Q}^{J)}\delta N\,.
\label{delta2XIJ}
\eeq
From the explicit expression for $\H$ in Eq.~(\ref{Fdef}), 
we immediately see that its antisymmetric structure 
 implies that 
${\H}_{,K}={\H}_{,KL}=0$.  We also find
\ba
{\H}_{<IJ>}&=&-2fX \perp_{IJ},
\\
 \H_{<IJ>,<KL>}&=&2f c_s^2\left(\tG_{I(K} \tG_{L)J}-\tG_{IJ}\tG_{KL}\right) ,
\\
 \H_{<IJ>,K}&=&  -2f X\left( G_{IJ,K}+G_{IJ}e^Le^M G_{LM,K}-e_Ie^LG_{JL,K}-e_Je^LG_{IL,K} 
\right) -2  f_{,K} X\perp_{IJ},
\ea
where the first two identities can be deduced from Eqs.~(\ref{PIJ}) and (\ref{P(IJ)(KL)}).
This readily gives
\beq
\label{delta}
\delta^{(1)}\tX=\delta^{(1)} X, \quad \delta^{(2)}\tX=\delta^{(2)}X +{fX}\perp_{IJ} h^{ij} \partial_i Q^I\partial_j Q^J.
\eeq

\section*{Appendix 3: third-order action}
\label{App:3}

The following relations are useful for determining the third-order action:
\be
X^{IJ}= \frac{1}{2}\dot{\phi}^I \dot{\phi}^J + \delta^{(1)}X^{IJ} +  \delta^{(2)}X^{IJ} +  \delta^{(3)}X^{IJ}
\ee
where we have used Eq.~(\ref{NA}) to rewrite (\ref{delta1XIJ}) and (\ref{delta2XIJ}) in the form
\ba
 \delta^{(1)}X^{IJ} &=& -2\N_A  X^{IJ} Q^A+ \dot{\phi}^{(I} \dot{Q}^{J)}
\label{delta1X}
\\
 \delta^{(2)}X^{IJ} &=& -2\N_A  \dot{\phi}^{(I} \dot{Q}^{J)} Q^A + 3 \N_A \N_B X^{IJ} Q^A Q^B 
  + \frac{1}{2}\dot{Q}^I \dot{Q}^J
- \dot{\phi}^{(I} \partial_iQ^{J)} (\delta N^i)
- \frac{1}{2}h^{ij}\partial_iQ^I \partial_jQ^J 
\label{delta2X} \, .
\ea
We also have
\ba
P_{<IJ>} \delta^{(3)}X^{IJ} &=& -4P_{<IJ>}X^{IJ} \N_A \N_B \N_C Q^A Q^B Q^C + 6 H \N_A \N_B \N_C Q^A Q^B \dot{Q}^C
 \nn
 \\
 && - P_{<BC>} \N_A Q^A \dot{Q}^B \dot{Q}^C
+ 4 H \N_B \N_A Q^A \partial_iQ^{B} \delta N^i
\nn
 \\
 &&- P_{<AB>} \dot{Q}^A \partial_iQ^{B} \delta N^i
\ea
as well as
\ba
\delta^{(2)}P &=& P_{<IJ>} \delta^{(2)}X^{IJ} + \frac{1}{2} P_{<IJ>,<KL>} \delta^{(1)}X^{IJ}  \delta^{(1)}X^{KL} + P_{<IJ>,K}Q^K\delta^{(1)}X^{IJ} 
 + \frac{1}{2} P_{,KL}Q^K Q^L
\ea
and
\ba
\delta^{(3)}P &=& P_{<IJ>} \delta^{(3)}X^{IJ} + \left[ P_{<IJ>,<KL>} \delta^{(2)}X^{IJ} \delta^{(1)}X^{KL} + P_{<IJ>,K}Q^K \delta^{(2)}X^{IJ} \right]
\nn
\\
&& 
+ \frac{1}{6} \left[ P_{<IJ>,<KL>,<MN>} \delta^{(1)}X^{IJ} \delta^{(1)}X^{KL} \delta^{(1)}X^{MN} + P_{,IJK}Q^IQ^JQ^K \right]
\nn
\\
&&+ \frac{1}{2} \left[ P_{<IJ>,<KL>,M} Q^M \delta^{(1)}X^{IJ} \delta^{(1)}X^{KL}
+ P_{<MN>,IJ} Q^IQ^J \delta^{(1)}X^{MN} \right].
\ea


\end{document}